\documentclass[a4paper]{article}

\usepackage[margin=1in]{geometry} 

\usepackage{amsmath}
\usepackage{amsthm}
\usepackage{amssymb}
\usepackage{multirow}
\usepackage{booktabs}
\usepackage[utf8]{inputenc}
\usepackage{hyperref}
\hypersetup{
	unicode,
	pdfauthor={Author One, Author Two, Author Three},
	pdftitle={A simple article template},
	pdfsubject={A simple article template},
	pdfkeywords={article, template, simple},
	pdfproducer={LaTeX},
	pdfcreator={pdflatex}
}

\usepackage[sort&compress,numbers,square]{natbib}
\theoremstyle{plain}

\theoremstyle{definition}

\usepackage{graphicx, color}
\graphicspath{{fig/}}

\usepackage{algorithm, algpseudocode} 
\usepackage{mathrsfs} 

\usepackage{lipsum}

\title{Evaluation of Reliability Criteria for News Publishers\\ with Large Language Models}

\author{Manuel Pratelli$^{1,2}$ \and John Bianchi$^{2}$ \and Fabio Pinelli$^{2}$ \and Marinella Petrocchi$^{1,2}$}

\date{
	$^1$IIT-CNR, Pisa \\ \texttt{\{manuel.pratelli, marinella.petrocchi\}@iit.cnr.it}\\%
	$^2$IMT School for Advanced Studies Lucca\\ \texttt{\{johnbianchi, fabiopinelli\}@imtlucca.it}\\[2ex]%
}

\begin{document}
	\maketitle
	
\begin{abstract}
    In this study, we investigate the use of a large language model to assist in the evaluation of the reliability of the vast number of existing online news publishers, addressing the impracticality of relying solely on human expert annotators for this task.
    In the context of the Italian news media market, we first task the model with evaluating expert-designed reliability criteria using a representative sample of news articles. We then compare the model's answers with those of human experts.
    The dataset consists of 340 news articles, each annotated by two human experts and the LLM. 
    Six criteria are taken into account, for a total of 6,120 annotations. 
    We observe good agreement between LLM and human annotators in three of the six evaluated criteria, including the critical ability to detect instances where a text negatively targets an entity or individual. For two additional criteria, such as the detection of sensational language and the recognition of bias in news content, LLMs generate fair annotations, albeit with certain trade-offs. Furthermore, we show that the LLM is able to help resolve disagreements among human experts, especially in tasks such as identifying cases of negative targeting.
 
\end{abstract}

	\section{Introduction}
The decline of the traditional journalism system, combined with difficulties in maintaining editorial control, has raised concerns about the quality of information disseminated by online news publishers. Specialized organizations~\cite{newsguard,mediabiasfactcheck,iffyindex,globaldisinformationindex,adfontesmedia}, have been instrumental in assessing the reliability of many online news sources. 

\begin{figure}[ht!]
\centering
    \includegraphics[width=.6\linewidth]{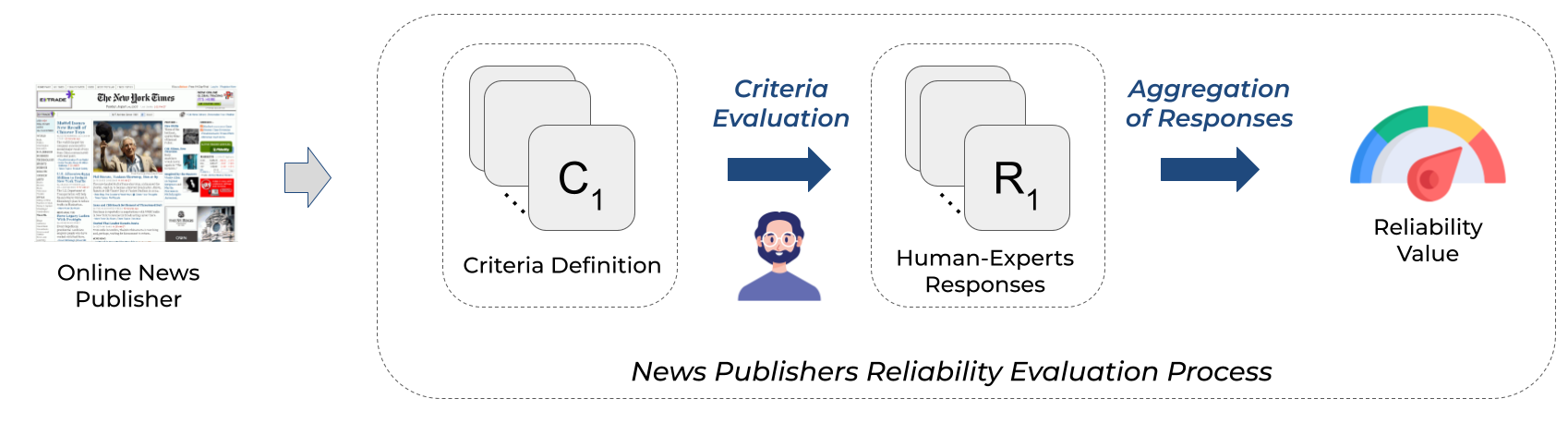} 
    \caption{Traditional Approach to Evaluating the Reliability of a News Publisher} 
    \label{fig:pub_trust_eval}
\end{figure}

Each organisation conducts the evaluation process by defining a specific set of criteria and identifying the expert profiles required for their assessment (e.g., journalists or trained annotators). Criteria typically consider elements like the inclination to publish propaganda or politically biased content~\cite{bazmi2023multi,kim2019combating}. 
The phase of the criteria evaluation process generally lacks automation.
This process can be modeled as depicted in Figure \ref{fig:pub_trust_eval}. 
The organization in charge of evaluating publishers sets criteria for "good journalism" and assigns experts in the field to manually evaluate the publisher against these criteria.
The evaluation results are aggregated at the criterion level, taking into account potential disagreements among experts, to produce a final score - often numerical in nature. This score serves as a holistic measure of the publisher's overall reliability, providing a comprehensive assessment without detailing the specific contributions of individual criteria.
Although these scores offer valuable insights \cite{celadin2023displaying,prike2024source,kim2019combating,celadin2023displaying}, the process of evaluating individual news outlets is labor-intensive and time-consuming. Manual assessments, where expert annotators scrutinize ownership information and content, remain crucial yet demanding~\cite{Freeze2021,doi:10.1287/mnsc.2019.3478}. 

In recent years, Large Language Models (LLMs) have excelled at generating text that closely mimics human language \cite{doi:10.1073/pnas.2208839120}. Their capabilities extend to various natural language processing tasks, such as sentiment analysis and text summarization \cite{ye2023comprehensive}. This capability has paved the way for numerous potential applications, including improving educational processes and providing answers to medical questions \cite{10.1001/jamainternmed.2023.1838}. 

On the one hand, we have an accurate but time-consuming evaluation process: it takes months to evaluate a single news source, starting with the selection and training of annotators, through the actual evaluation of specific criteria, and ending with the calculation of the final score, not to mention that after months the situation of the media outlet itself may have changed and require a new analysis. 
On the other hand, a generative intelligence whose evolution seems to progress constantly, see, e.g., the recent release of GPT-4o (omni)\cite{openai}.  

\begin{figure}[ht!]
\centering  \includegraphics[width=\linewidth]{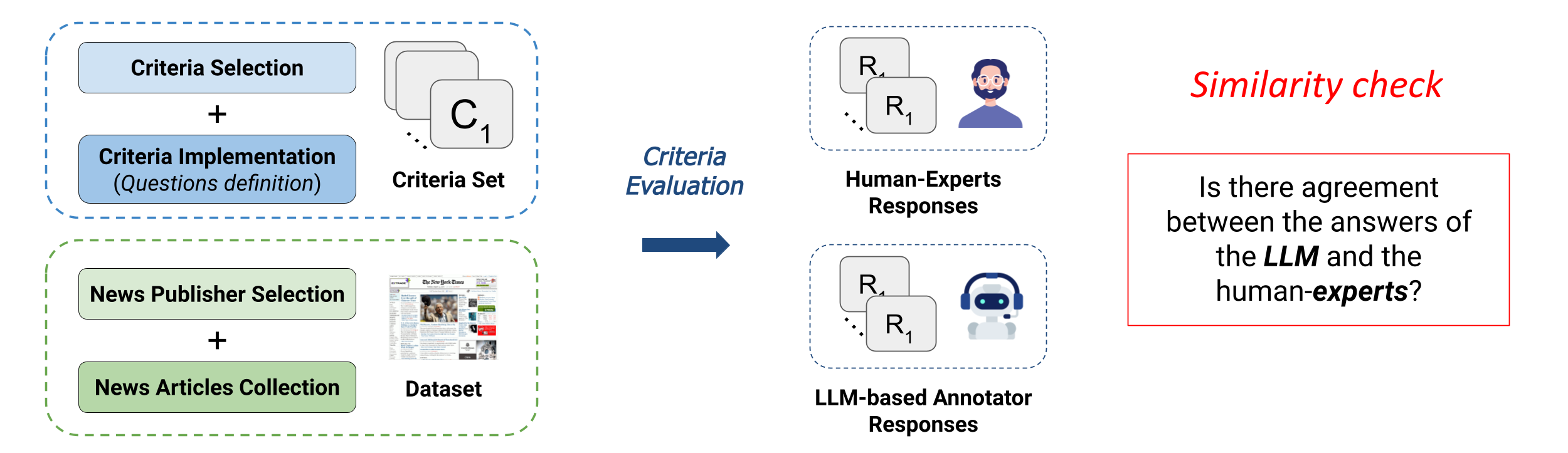} 
    \caption{Approach} 
    \label{fig:core_idea}
\end{figure}

\indent\textbf{Objective and Approach.}
The main goal of this work is to develop a more scalable and more efficient trust evaluation process by incorporating automation into the traditional approach of manually evaluating good journalism criteria defined by specialized organizations, see Figure~\ref{fig:pub_trust_eval}.
 To the best of our knowledge, this is the first study that proposes to introduce automation into the task of evaluating such criteria. 
 To conduct the experiment, as shown in Figure \ref{fig:core_idea}, we use a two-step process. In the first step, we perform the two following tasks in parallel: (i) the selection and implementation of a set of quality criteria, and (ii) the collection of a representative sample of publishers and news articles. 
In the second step, three experienced human annotators 
evaluate the criteria on each of the sampled articles. Their answers are compared with those of an LLM. The intuition underlying this study is that agreement between human experts and the LLM on individual criteria leaves room for automation of related subtasks. In addition, we explore the potential of the LLM to assist in resolving disagreements among human experts.


\indent\textbf{Research Questions.} 
Achieving our goal translates into answering the following questions:
\begin{itemize}
    \item \textbf{RQ1:} How to define quality prompts to help LLMs correctly evaluate good journalism criteria to assess the reliability level of a publisher?
    \item \textbf{RQ2:} 
    How good is an LLM at aligning with the responses given by human-experts when evaluating the good journalism criteria?
    \item \textbf{RQ3:} What are the types of questions for which an LLM can effectively support, dare we say even replace, an expert annotator in the process of evaluating the reliability of a publisher?
\end{itemize}


\indent\textbf{Results.}
Through a process of refinement, we are able to write prompts that are optimized so that 1) the LLM understands the questions it needs to answer, and 2) the answers are consistent with those of human annotators.
The results show agreement between the LLM and the annotators in recognizing the type of the news, the presence of a fact-based \textit{lede} at the beginning of the article and whether the text negatively targets someone or something. Also, the evaluation of labor-intensive criteria, such as checking for the presence 
of article bias and 
sensational language, achieves a fair level  of agreement.
In addition, the LLM can provide valuable support in cases of disagreement between human experts in different evaluation tasks.


\indent\textbf{Applications.} The ability to automatically generate high-quality, criteria-specific responses from the analysis of news articles opens up at least two potential avenues of application:
\begin{itemize}
\item On an organizational level, automated evaluation of criteria can be integrated into existing workflows for assessing the reliability of news publishers. This is particularly effective in resource-constrained environments where automation provides scalability. For example, assessing labor-intensive criteria on a large volume of news articles is often not feasible due to limited expertise or funding. Automation, on the other hand, enables large-scale, near-real-time monitoring and analysis, significantly improving the efficiency and reach of reliability assessments.
    \item At the user level, criteria-specific responses can increase readers' awareness of the content they consume in real time. The goal is to highlight potential deviations in specific aspects (e.g., sensational language) from established quality standards of good journalism. This approach translates expert practices into tools accessible to end-users, fostering critical thinking and enabling more informed and reflective engagement with news content.
\end{itemize}

\section{Related Work}


\paragraph{Towards Automating the Evaluation of News Source Reliability: Challenges, Advances, and Gaps} Several studies have proposed lists of criteria to distinguish reliable news sites from unreliable ones. Notable examples include criteria developed by: (i) independent organizations, such as the Global Disinformation Index (GDI) \cite{globaldisinformationindex}, NewsGuard \cite{newsguard}, and Media Bias Fact Check (MBFC) \cite{mediabiasfactcheck};
(ii) platforms, such as Facebook's "Tips to Spot False News" \cite{fbtips};
(iii) academic research \cite{heuer2024reliability,zhang2018structured}; and
(iv) collaborative projects \cite{trustprojindicators,edmotraining}.

The criteria are designed for different applications and annotator profiles, ranging from lay users to expert human annotators. As a result, the granularity of the evaluation and the complexity of the tasks required vary. For example, Facebook's  guidelines \cite{fbtips} provide simple tips to help ordinary users identify misinformation and make informed decisions. In contrast, the criteria developed by GDI \cite{globaldisinformationindex} and NewsGuard \cite{newsguardcriteria} are tailored for expert annotators - such as trained professionals or journalists. While these criteria are clear and interpretable, applying them to a news article often requires expertise in the field. In other words, experts and non-experts (e.g., crowd raters) might disagree \cite{bhuiyan2020investigating}. For example, verifying the presence of bias in the textual content of news articles might not be trivial for a common user to evaluate.

In recent years, substantial evidence has emerged supporting the effectiveness of professional and user-generated ratings applied at the text level \cite{celadin2023displaying} or source level \cite{celadin2023displaying,prike2024source,kim2019combating} in mitigating the spread of low-quality content. In particular, ratings based on expert criteria show stronger effects in reducing the intention to share low-quality news \cite{celadin2023displaying}.
However, manually evaluating content at the professional level, such as fact-checking individual news headlines \cite{celadin2023displaying}, poses significant scalability challenges. Even more, continuous evaluation of source-level ratings, which often aggregate ratings of article-level criteria \cite{newsguardcriteria,globaldisinformationindex}, is highly impractical given the massive volume and rapid distribution of news content online.

Given the challenges of scalability and the recognition of source rating estimation as a key intervention to mitigate the spread of low-quality content \cite{kozyreva2024toolbox}, we highlight the paucity of research addressing these critical needs.

Some recent work aims to automate the evaluation of the reliability of news sources, exploring different approaches. One method relies solely on user interactions within social media discussions, ignoring the content of the articles themselves \cite{pratelli2024unveiling}. Another approach infers the reliability of the news publisher by classifying the content of individual articles \cite{bianchi2024evaluating}. A third line of research involves using LLMs to directly evaluate news outlets \cite{MenczerGPT2023}.

With specialized organizations in mind, the Global Disinformation Index (GDI) uses LLMs to encode tagged phrases from known disinformation content, creating digital filters. These filters are applied to new material to flag sites for manual review if a significant number of sentences match established disinformation patterns. A team of trained analysts conducts weekly manual reviews \cite{gdiproduct}. To our knowledge, GDI does not automate the process of criteria evaluation. Instead, it uses LLM to assess similarity with analyst-generated narrative filters that serve as examples of disinformation themes.
NewsGuard uses artificial intelligence in proprietary machine learning processes to identify false claims circulating online in real time \cite{newsguardsafetysuite}.

The closest work to ours that we found by studying the literature is that of Yang and Menczer \cite{MenczerGPT2023}. It considers more than 7,000 news outlets, in a variety of languages, and asks ChatGPT 3.5 (March 1, 2023 release) to rate each of these news outlets from 0 to 1 (0 no credibility, 1 max credibility), giving as input only the name of the news outlet. The result was then compared with the ratings given to them NewsGuard and the Media Bias Fact Check, as well as by aggregate human ratings produced by Lin and others \cite{10.1093/pnasnexus/pgad286}. The ratings given by ChatGPT correlate with those of the expert annotators (Spearmam’s $\rho$ = 0.54, p < 0.001). 
Although one of the goals of the present work is similar to that of \cite{MenczerGPT2023}, i.e., to evaluate the agreement between the LLM and expert annotators regarding the reliability of online news outlets, the two studies differ in methodology. In terms of methodology, Yang and Menczer's prompt is concise and high-level, asking the GPT to rate the reliability of the media outlet within a range of values. Our prompt is more complex: we ask the LLM to rate 6 criteria out of 340 news articles, and we compare all LLM responses with those of human annotators. 
The capillarity of the study (asking the LLM a series of specific questions) is preparatory to understanding which steps in the process can be automated through the use of LLMs.
Thus, to the best of our knowledge, no prior work has proposed a methodology for using LLMs to automate the evaluation of a complete set of widely accepted, expert-designed criteria for assessing the quality of online information. 

We identify two key gaps in the current scientific literature: (i) the lack of a generally accepted list of evaluation criteria, and (ii) the lack of a comprehensive ground truth dataset to support research in this area.
Focusing on the Italian media landscape, this study represents the first attempt to address these gaps. Specifically, we: (i) analyze a set of "worthy-to-automate" criteria: we select them because they are adopted for conducting real-world news website reliability evaluation from two expert-designed frameworks: NewsGuard \cite{newsguardcriteria} and the Global Disinformation Index (GDI) (see Table \ref{tab:GDI_criteria});
 (ii) explore the potential 
of LLMs to automate their evaluation; and (iii) provide a resulting annotated dataset to facilitate further research in this area.

\section{Methodology}
 
This section presents our methodology. The goal is to define appropriate prompts for the LLM to evaluate journalistic criteria used to assess the reliability of online publishers, and measure the agreement between the LLM's answers and those of experienced annotators.



\subsection{Framework overview}


We start with the selection (Section \ref{sec:criteria_selection}) and implementation (Section \ref{sec:criteria_implementation}) of a set of criteria for evaluating the reliability of online news publishers (upper left blue dotted box in 
Figure \ref{fig:core_idea}). 
As explained in \ref{sec:criteria_implementation}, in the context of this study, criteria implementation means defining a list of questions that can be evaluated by both a human expert and an LLM-based annotator. Each question aims to highlight a specific aspect of a news item according to the corresponding criterion.
 Second, as described in Section \ref{sec:dataset}, we proceed to select a set of publishers to evaluate and extract a sample of news articles from those publishers 
(lower left green dotted box in 
Figure \ref{fig:core_idea}). Then, in \ref{sec:criteria_eval}, we ask three experienced annotators and the LLM to answer the questions representing the criteria. Finally, in the \textit{Results} section, we evaluate the agreement between the expert answers and the LLM answer (right red box in 
Figure \ref{fig:core_idea}). In cases where there is disagreement among experts, we explore the potential of the LLM to facilitate resolution of these conflicts.


\subsection{Criteria Selection}\label{sec:criteria_selection}

The goal of this section is to select a \textit{worthy-to-automate} list of criteria. As mentioned earlier, these criteria must be both widely recognized and labor intensive to evaluate manually.

The starting point for this analysis is the list of criteria proposed by \cite{pratelli2022structured}. This work identifies, from the complete list of GDI criteria \cite{mediamarketItaly2022}, a preliminary set of \textit{worthy-to-automate} criteria. 
A detailed explanation of these criteria can be found in Table \ref{tab:GDI_criteria}. These criteria are selected because,  although they have different names, and their definitions do not match NewsGuard's definition, there is a strong conceptual correspondence between them \cite{pratelli2022structured}.
The NewsGuard list of criteria is particularly noteworthy, as it is widely used \cite{heuer2024reliability}) too. This consistency in definition across both organizations assures that these criteria are widely accepted as effective for conducting real-world reliability assessments of news publishers.

 The tests in \cite{pratelli2022structured} were conducted on the Italian media market, which is also the focus of this study. For the scope of this paper, however, we have excluded some criteria from the original set. In particular, we will not consider in our analysis the criteria that do not involve the analysis of a written text (like \textit{VisPres}) and the (trivial-to-evaluate) features that can be extracted using heuristic methods (\textit{ByInfo}). Also, \textit{Common Coverage} is excluded due to their reliance on multi-source analysis. The last six criteria in Table \ref{tab:GDI_criteria} cover broader editorial and operational aspects of news outlets, such as editorial independence, funding sources, and ownership structure. These require access to meta-information that is not typically part of the article text, making them less amenable to automated evaluation by an LLM.

The resulting set of \textit{worthy-to-automate} criteria is as follows:
\begin{itemize}
    \item \textit{Headline Accuracy}: Assesses the congruence between a headline and the corresponding article content.
    \item \textit{Lede Presence}: Checks for the presence of a factual summary at the beginning of an article.
    \item \textit{Negative Targeting}: Evaluates whether the article negatively targets specific individuals or institutions.
    \item \textit{Article Bias}: Gauges the fairness and balance of the article's content.
    \item \textit{Sensational Language}: Identifies emotionally charged or exaggerated language that could mislead readers.
\end{itemize}



\begin{table*}[ht!]
\tiny
\caption{Criteria names and definitions}
\label{tab:GDI_criteria}
\resizebox{\linewidth}{!}{
\begin{tabular}{ p{0.18\linewidth} p{0.12\linewidth} p{0.7\linewidth} }
\hline
\textbf{Criteria Name} & \textbf{Short name}  & \textbf{Definition} \\ \hline
Headline accuracy & HeadAcc & Rating for how accurately the story’s headline describes the content of the story. Indicative of clickbait \\ \hline
Byline information & Bylnfo & Rating for how much information is provided in the article’s byline. Attribution of stories creates accountability for their veracity \\ \hline
Lede present & LedePres & Rating for whether the article begins with a fact-based lede. Indicative of fact-based reporting and high journalistic standards \\  \hline
Common coverage & ComCov & Rating for whether the same event has been covered by at least one other reliable local media outlet. Indicative of a true and significant event \\  \hline
Recent coverage & RecCov & Rating for whether the story covers a news event or development that occurred within 30 days prior to the article’s publication date. Indicative of a newsworthy event, rather than one which has been taken out of context \\  \hline
Negative targeting & NegTarg & Rating for whether the story negatively targets a specific individual or group. Indicative of hate speech, bias or an adversarial narrative. \\  \hline
Article bias & ArtBias & Rating for the degree of bias in the article. Indicative of neutral fact-based reporting or well-rounded analysis. \\  \hline
Sensational language & SensLang & Rating for the degree of sensationalism in the article. Indicative of neutral fact-based reporting or well-rounded analysis. \\  \hline
Visual presentation & VisPres & Rating for the degree of sensationalism in the visual presentation of the article. Indicative of neutral fact-based reporting or well-rounded analysis. \\  \hline
Attribution & Attr & Rating for the number of policies and practices identified on the site. Assesses policies regarding the attribution of stories, facts, and media (either publicly or anonymously); indicative of policies that ensure accurate facts, authentic media and accountability for stories. \\  \hline
\multirow{2}{*}{Comment policies} & \multirow{2}{*}{CommPol} & Rating for the number of policies identified on the site. Assesses policies to reduce disinformation in user-generated content. Rating for the mechanisms to enforce comment policies identified on the site. Assesses the mechanism to enforce policies to reduce disinformation in user-generated content\\
\hline
\multirow{4}{*}{\shortstack{Editorial principles\\ and practices}} & \multirow{4}{*}{EdPrincPract} &  Rating for the number of policies identified on the site. Assesses the degree of editorial independence and the policies in place to mitigate conflicts of interest \\
 & &  Rating for the degree to which the site is likely to adhere to an ideological affiliation, based on its published editorial positions. Indicative of politicised or ideological editorial decision-making \\  
\hline
\multirow{2}{*}{Ensuring accuracy} & \multirow{2}{*}{EnsAcc} &  Rating for the number of policies and practices identified on the site. Assesses policies to ensure that only accurate information is reported \\
\multirow{3}{*}{Funding} & \multirow{3}{*}{Fund} &  Rating for the number of revenue sources identified on the site. Indicative of possible conflicts of interest stemming for over-reliance on one or few sources of revenue \\
\hline
\multirow{2}{*}{Ownership} & \multirow{2}{*}{Own} &  Rating based on the number of distinct executive- or board-level financial and editorial decision-makers listed on the site. Indicative of a separation between financial and editorial decision-making, to avoid conflicts of interest. \\
 \hline
\end{tabular}
}
\end{table*}

\subsection{Criteria Implementation}\label{sec:criteria_implementation}
In this paper, implementing a criterion means defining a clear and concise question, along with potential responses, that can be effectively evaluated by both human experts and LLM-based annotators, using the text of a news article as input. Table \ref{tab:selected_GDI_criteria_and_iplementation} details how we have implemented the criteria selected in Section \ref{sec:criteria_selection}. Most questions closely align with the descriptions of the criteria provided by GDI. For example, GDI criterion \textit{Negative Targeting}: ``\textit{Rating of whether the story negatively targets a specific individual or group}'' is made concrete by the question: ``\textit{Does the article negatively target individuals or groups?}''
On the same prompt, we also ask the model to indicate
what issue the individual or group is negatively targeted on, to assess its ability to identify specific groups or minorities that are often negatively targeted in the news. We also introduce a meta-criterion (\textit{Type}) to assess whether the LLM can accurately determine the type of article it is analyzing, such as distinguishing between straight news and other journalistic forms, like editorials, satire, or soft news \cite{diffnewsinvestig}.



\begin{table*}[th]
    \small
    \caption{Selected GDI criteria and their implementation}
    \label{tab:selected_GDI_criteria_and_iplementation}
    \resizebox{\textwidth}{!}{
        \begin{tabular}{p{2.4cm}p{1.3cm}p{7.2cm}p{5.2cm}}
            \toprule
            Criteria Name & Short name & Implementation & Answers' Options \\
            \midrule
            Headline accuracy  & HeadAcc& How accurate is the news’s headline with the content of the news? & Inaccurate/Quite inaccurate/Quite accurate/Accurate\\
            Lede present & LedePres & Does the article start with a summary of the main facts? & Yes/No \\ 
            Negative targeting & NegTarg & Does the article negatively target individuals or groups? Indicate what issue the group or individual is negatively targeted on. & Yes/No and issue ( Politics / Gender / Religion / Other )\\  
            Article bias & ArtBias & How much biased is the article? & Biased/Quite biased/Quite unbiased/Unbiased\\  
            Sensational language & SensLang & How sensational is the tone of the news? & Sensational/Quite sensational/Quite neutral/Neutral \\ 
    
            \midrule
            \textbf{Meta-criterion} &&&\\
            Type of news & Type & What kind of news are you reading? & Straight news/Editorial/Investigation/Satire/Soft News\\
    
    

            \bottomrule
        \end{tabular}
    }
\end{table*}

\begin{table*}[!ht]
    \scriptsize
    \centering
    \caption{Online publishers considered in this study (source: \cite{mediamarketItaly2022}) }
    \resizebox{\textwidth}{!}{
        \begin{tabular}{llllll}
        \toprule
             Publisher & Website & Publisher & Website & Publisher & Website \\ 
             \midrule
            Avvenire & www.avvenire.it & Il Corriere Del Giorno & www.ilcorrieredelgiorno.it & Il Piccolo & www.ilpiccolo.gelocal.it \\
            Corriere Della Sera & www.corriere.it & Il Fatto Quotidiano & www.ilfattoquotidiano.it & Il Post & www.ilpost.it \\
            Domani & www.editorialedomani.it & Il Foglio & www.ilfoglio.it & Il Primato Nazionale & www.ilprimatonazionale.it \\
            Il Gazzettino & www.ilgazzettino.it & Il Giornale & www.ilgiornale.it & Il Quotidiano Del Molise & www.quotidianomolise.com \\
            Il Giornale Di Sicilia & www.gds.it & Il Giorno & www.ilgiorno.it & Il Resto Del Carlino & www.ilrestodelcarlino.it \\
            Il Giorno & www.ilgiorno.it & Il Manifesto & www.ilmanifesto.it & Il Secolo D'italia & www.secoloditalia.it \\
            Il Mattino & www.ilmattino.it & Il Messaggero & www.ilmessaggero.it & Il Sole 24 Ore & www.ilsole24ore.com \\
            Il Tirreno & www.iltirreno.gelocal.it & La Gazzetta Del Mezzogiorno & www.lagazzettadelmezzogiorno.it & La Nuova Ferrara & www.lanuovaferrara.gelocal.it \\
            La Nazione & www.lanazione.it & La Nuova Padania & www.lanuovapadania.it & La Nuova Sardegna & www.lanuovasardegna.it \\
            La Repubblica & www.repubblica.it & La Stampa & www.lastampa.it & La Veritá & www.laverita.info \\
            Libero & www.liberoquotidiano.it & Libertá & www.liberta.it & L'unione Sarda & www.unionesarda.it \\
            Open & www.open.online & Stopcensura & www.stopcensura.online &  &  \\
            \bottomrule
        \end{tabular}
    }
    \label{tab:informationSites}
\end{table*}

\subsection{Articles Dataset}\label{sec:dataset}
This study focuses on the Italian media market. To ensure an unbiased analysis, it is essential to gather textual data from articles that best represent the Italian online media landscape. The approach adopted involves collecting articles from a list of representative Italian news publishers. Specifically, a past study about the Italian online news market \cite{mediamarketItaly2022} proposed a balanced list of Italian news publishers, considering: (i) the media outlet distribution (national or local), (ii) the geographic location of the media outlet, (iii) the political orientation, and (iv) the disinformation risk scores. Table \ref{tab:informationSites} shows the online publishers selected in \cite{mediamarketItaly2022}. We have included these 34 publishers in this study because they have already been selected to be representative of a country's news market.

Furthermore, for the current analysis we have collected 
 news articles that were published at the same time as the articles in the original study. This is because the quality of a newspaper can vary over time, and we wanted to try to replicate the same distribution of news articles, both by source  and risk of exposure to disinformation.
Thus, for each of the 34 news publishers, we retrieved 10 articles published within a 7-month period, from April to October 2021, consistent with the timeframe and sample size used in the GDI analysis.

We would like to emphasize that the primary goal of this study is not to evaluate the reliability of a single publisher. Instead, our focus is on assessing the ability of a large language model to evaluate specific journalistic quality criteria applied to a curated collection of article texts. Thus, we had 340 articles evaluated by three human annotators and the LLM, making them answer 6 criteria per article, resulting in a total of 6,120 annotations.

The article texts were collected using an automated pipeline combining several tools. First, we employed the Selenium library \cite{selenium} to gather URLs from publishers’ websites. Next, we utilized the GNU Wget command \cite{wget} to download each article's HTML page. Finally, XPATH queries were used to extract the textual content, including the title and main text of each article.

\begin{table*}
    \small
    \caption{Criteria implementation refinement 
     }
    \label{tab:gpt_adapted_selected_criteria_implementation}
    \resizebox{\textwidth}{!}{
        \begin{tabular}{llll}
            \toprule
            Short Name & Question-Level Modification & Response-Level Modification & Description \\
            \midrule
            HeadAcc & Rephrased for clarity & None & - \\
            LedePres & Rephrased for clarity & None & Introduced the definition of lede into the question \\
            NegTarg & None & None & - \\
            ArtBias & None & Response options simplified to Biased/Unbiased & - \\
            SensLang & None & Response options simplified to Sens/Neutral & - \\
    
            \midrule
            \textbf{Meta-criterion} &&& \\
            Type & None & Response option changed from "\textit{Editorial}" to "\textit{Opinion}" & - \\
            
            \bottomrule
        \end{tabular}
    }
\end{table*}
\subsection{Criteria Evaluation}\label{sec:criteria_eval}
This section introduces the human annotators and explains their annotation process. In addition, the section presents the design of the LLM-annotator prompts. 

\indent\textbf{Human Expert Annotations.} Initially, we sanitized the text of the articles to ensure anonymity of publishers and authors to prevent biasing the annotators' judgments. Then, we asked two experts from the Media and Communication Unit of our department, and a Ph.D. student in Data Science, to annotate the articles.
Articles, questions, and options for answers were uploaded to the Google Drive platform. Each article was annotated by two of the three annotators. 
The annotators were able to consult with each other when in doubt: This is because we want their annotation to be a ground truth against which we can later compare LLM responses.

\indent\textbf{Prompt Design.} The prompt design process ends with the formulation of the questions shown in Table \ref{tab:selected_GDI_criteria_and_iplementation}.
For each of the articles, we ask the LLM the questions listed in the table (column \textit{Implementation}) and give some options for the answers (column \textit{Answers' Options}). It is important to note that since our dataset consists of Italian-language news, the questions we pose to the LLM and the answers we expect are in Italian. In Table \ref{tab:selected_GDI_criteria_and_iplementation}, we have provided the English translation for the convenience of the reader.
The validity of the answers  is compared to the answers of the expert annotators, which we consider to be ground truth. The agreement, measured by Cohen's Kappa \cite{JacobCohen}, between the model's answers and those of the experts provides a measure of the quality of the prompts.

During the prompt design phase, we analyze the cases with the most disagreements and identify three main motivations:
(i) the prompt wording is not adequate for correct processing by the LLM, (ii) there are errors in the experts' annotations, and (iii) the LLM has difficulty interpreting certain contexts or questions correctly.

For prompts categorized under (i), we reviewed the initial versions of questions and answers until the latter were consistent with one of the answer options. Consistency was evaluated on a selected subset of news articles.

In modifying the prompts, both at the \textit{question} level and the \textit{answer} level, we preserved the semantics so that the validity of the answers given by human annotators remain unchanged, while improving the model's interpretive capabilities. Table \ref{tab:gpt_adapted_selected_criteria_implementation} shows how we refined the prompts. 


The \textit{HeadAcc} criterion was rephrased to enhance clarity. Similarly, the question for the \textit{LedePres} criterion was revised to explicitly incorporate the definition of an article's \textit{lede} within the question itself.

Some response options remain unchanged. Specifically, \textit{HeadAcc}, \textit{LedePres}, and \textit{NegTarg} continue to be binary responses. For \textit{ArtBias} and \textit{SensLang}, as detailed in Section \ref{sec:prompt_gain}, converting the original responses into binary formats enhances the consistency and reliability of the scoring process. Regarding the \textit{Type} meta-criterion, modifying the term \textit{"Editorial"} to \textit{"Opinion"} has proven to improve the quality of the responses.

After testing the new prompts on a subset of the news articles and manually checking the consistency of LLM's
responses, we repeated the experiment on the entire dataset. In the cases of \textit{ArtBias} and \textit{SensLang}, to implement the changes outlined in Table \ref{tab:gpt_adapted_selected_criteria_implementation}, we reduced the original four-class responses to two classes. This remapping allows us to maintain both datasets without the need for additional data collection.

During the prompt refinement process, we addressed the cases that fell under category (i). The improvements achieved through this process and the ability of the LLM to align with human-annotators are detailed in section \ref{sec:prompt_gain}.
For cases categorized under (ii) and (iii), the observed issues stemmed from either human annotation errors or the intrinsic limitations of the model. 
In Section \ref{sec:disagreement_comparison}, we investigate scenarios where the LLM can support in resolving disagreements between experts.

\noindent\textit{Collection of LLM annotations:} The model is instructed to assume the persona of an experienced journalist to ensure that the answers are appropriate for that role. Each question was asked to LLM three times, this was to check that it was not answering inconsistently, perhaps as a result of hallucinations. The answers given by LLM were always the same for the three rounds, thus, a single answer was produced for each article and criterion.


\section{Experimental Setup}\label{sec:exp_setup}

\textit{Agreement Metric: }
We use Cohen's Kappa~\cite{JacobCohen} to measure the agreement between annotators. 
The formula to calculate Cohen's Kappa is as \(k = (p_o - p_e) \div (1 - p_e)\) where \(p_o\) is the observed probability of agreement between the annotators and \(p_e\) is the expected probability of agreement if the annotators assigned the labels randomly~\cite{artstein2008inter}. We calculate this metric using the implementation available in the scikit-learn library\cite{scikitcohen}. As for the interpretation of values, we refer to the descriptions by Mary L. McHugh~\cite{mchugh2012interrater}, which are based on Jacob Cohen's original interpretations. A kappa value between 0 and 0.20 indicates no agreement among raters, beyond what would be expected by chance. Values from 0.21 to 0.39 represent minimal agreement. A weak level of agreement is observed for values between 0.40 and 0.59, while values between 0.60 and 0.79 indicate a moderate agreement. A strong agreement is denoted by kappa values from 0.80 to 0.90. Finally, values above 0.90 suggest an almost perfect agreement between the raters.

\noindent\textit{Model selection and configuration:} As LLM, we selected ChatGPT-4o, the most recent model released by OpenAI at the time of the experiments \cite{hellogpt4o}. We configure the model's hyperparameters to optimize the quality of its output. The temperature is set to zero because we need consistent and predictable responses. The \textit{Maximum Length} parameter has not been used because article lengths vary widely and a fixed limit would not be appropriate in all cases. Both \textit{Frequency Penalty} and \textit{Presence Penalty} are set to their default values (0) because avoiding repetition is not a priority for our task.
The \textit{Stop Sequences} parameter is set to \textit{None} because we do not need to stop text generation at a particular end sequence. 

\section{Results}
First, we see how well the LLM's responses match those of the human annotators, considering the original prompts in Table \ref{tab:selected_GDI_criteria_and_iplementation}. Then, in cases of strong disagreement, we see if using the refined prompts (Table \ref{tab:gpt_adapted_selected_criteria_implementation}) makes things better.
As a third analysis, we look at the cases where the human annotators disagree to see if the LLM could provide support.  


\subsection{Agreement between experts and LLM}\label{sec:prompt_gain}
\begin{figure}[ht!]
    \centering
    \includegraphics[width=.6\linewidth]{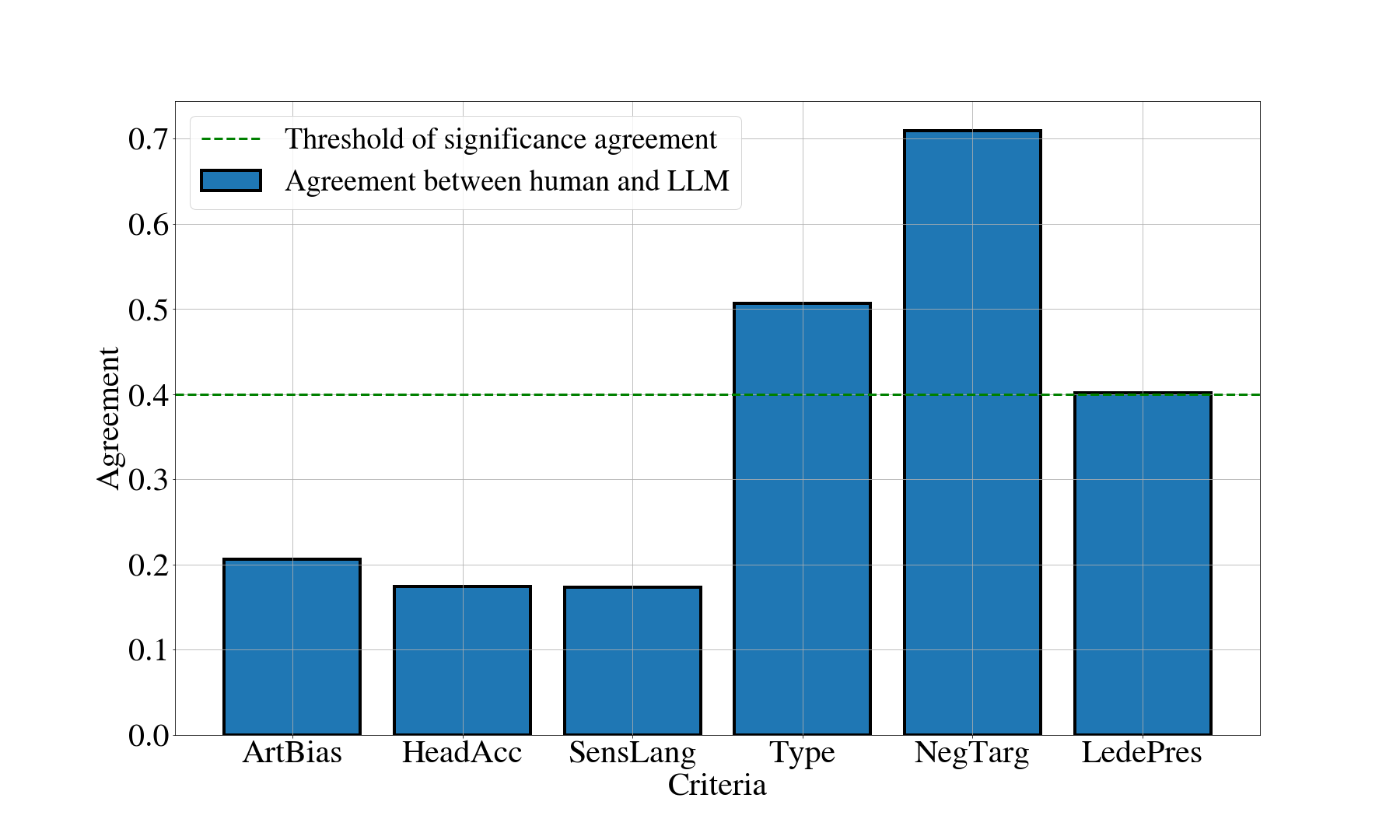} 
    \caption{Average agreement between experts and LLM. Note: only cases where experts agree between each other}  \label{fig:best_agreement}
\end{figure}
In Figure \ref{fig:best_agreement}, we report the agreement between LLM and human annotators \textit{for only those cases where human annotators agree with each other}. Essentially, for each article and criterion, the single annotation value determined by human experts (equal to the two concordant annotation values) is compared with the response provided by the LLM.

For 3 of the 6 criteria analyzed, the agreement between LLM and experts exceeds (or equals, in the case of LedePres) the minimum acceptance threshold set at 0.4. On the other hand, when assessing the degree of sensationalism in the article or the presence of bias, LLM disagrees with the expert annotators, who let us remember that it is the ground truth for us. Specifically, we obtain \(k\) = 0.2064 for \textit{ArtBias} and \(k\) = 0.1732 for \textit{SensLang}. We get the highest value of the agreement for \textit{NegTarg1}, \(k\) = 0.7089. Without prompt modification, the LLM is very careful to detect whether a news text is negatively targeting a group or a person. 

We went on to analyze in more detail what happens to the LLM when it has to evaluate the criteria \textit{SensLang} and  \textit{ArtBias}.

\begin{figure}[ht!]
    \centering
    \includegraphics[width=.6\linewidth]{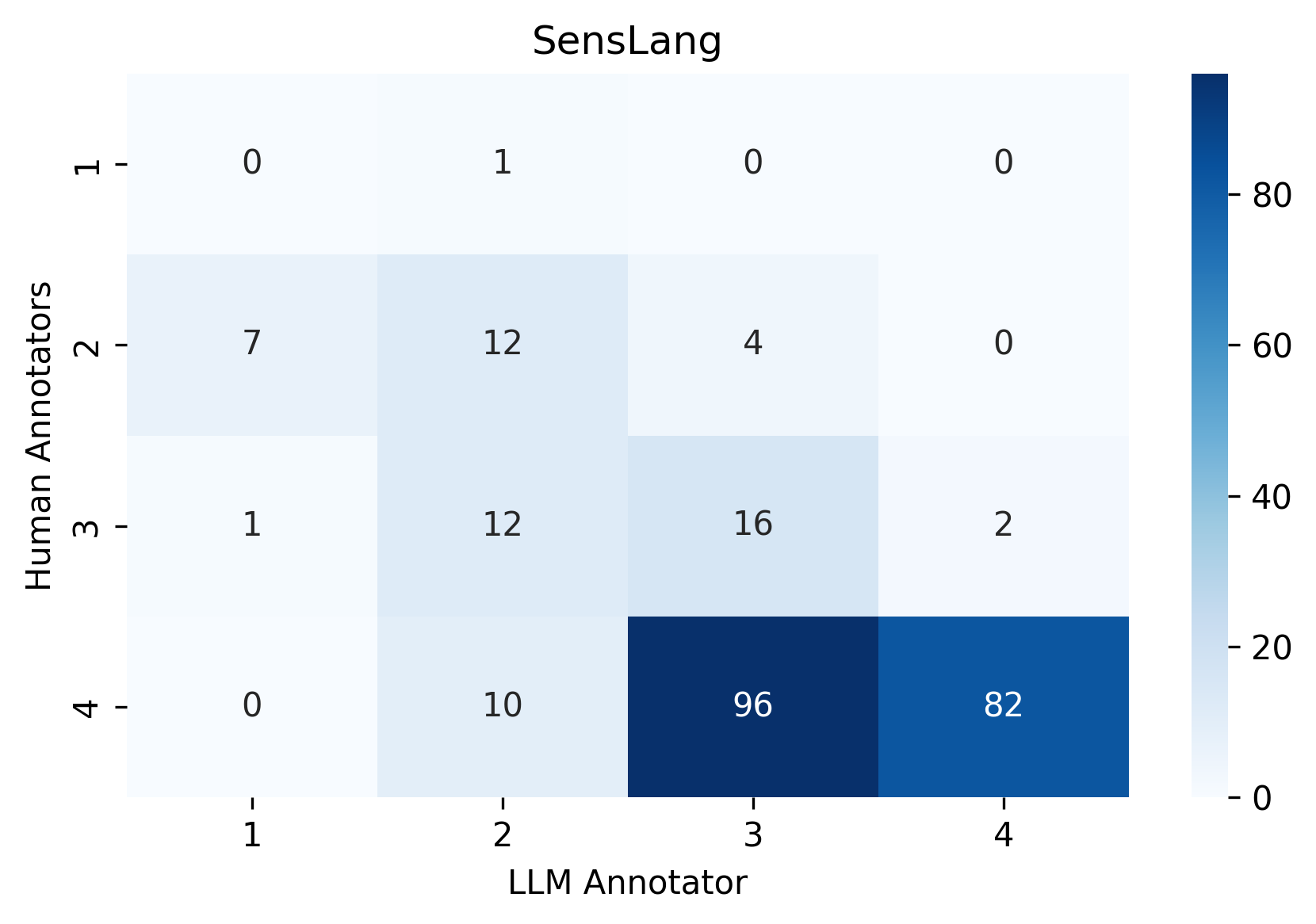} 
    \caption{Confusion Matrix for Criterion: Sensational Language. 1: Sensational; 4: Neutral}  \label{fig:cm_senslang}
\end{figure}

\begin{figure}[ht!]
    \centering
    \includegraphics[width=.6\linewidth]{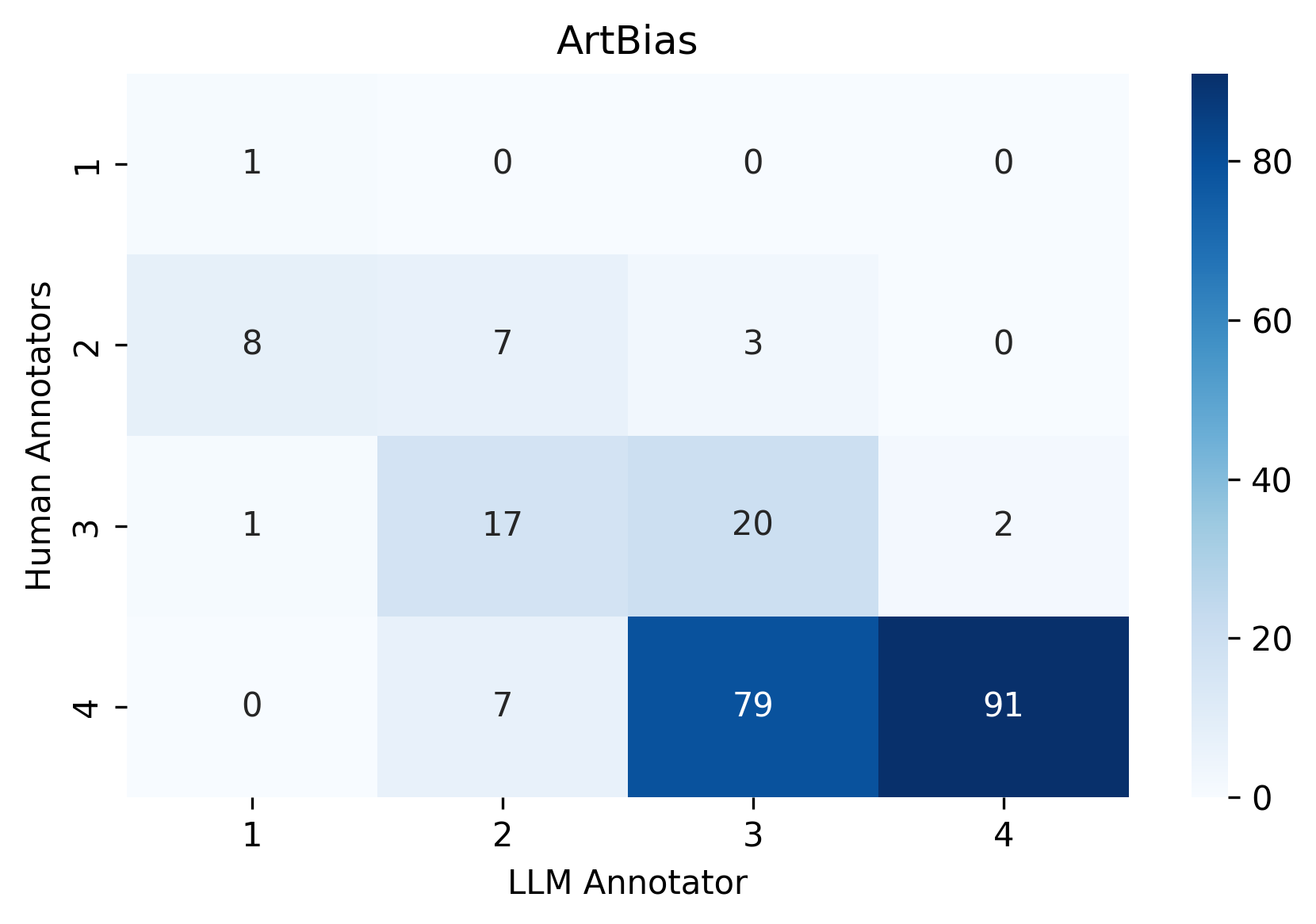} 
    \caption{Confusion Matrix for Criterion: Article Bias. 1: Biased; 4: Unbiased}  \label{fig:cm_artbias}
\end{figure}

Analyzing the confusion matrix of \textit{SensLang} (Figure \ref{fig:cm_senslang}) and \textit{
ArtBias} (Figure \ref{fig:cm_artbias}), we see that the LLM is more sensitive (i.e., it detects sensational language and bias where human annotators do not); however, it makes errors mainly in neighboring classes. In both confusion matrices, 1 means maximum degree of sensationalism/ bias, and 4 means no sensationalism/no bias. Looking at the case of sensationalism, for example, we see that for experts, the articles read are mostly neutral (10+96+82), while for LLM, only (82+2) have no trace of sensationalism. This is generally true for both matrices: LLM finds higher levels of sensationalism and bias than humans.

\begin{figure}[ht!]
    \centering
    \includegraphics[width=.5\linewidth]{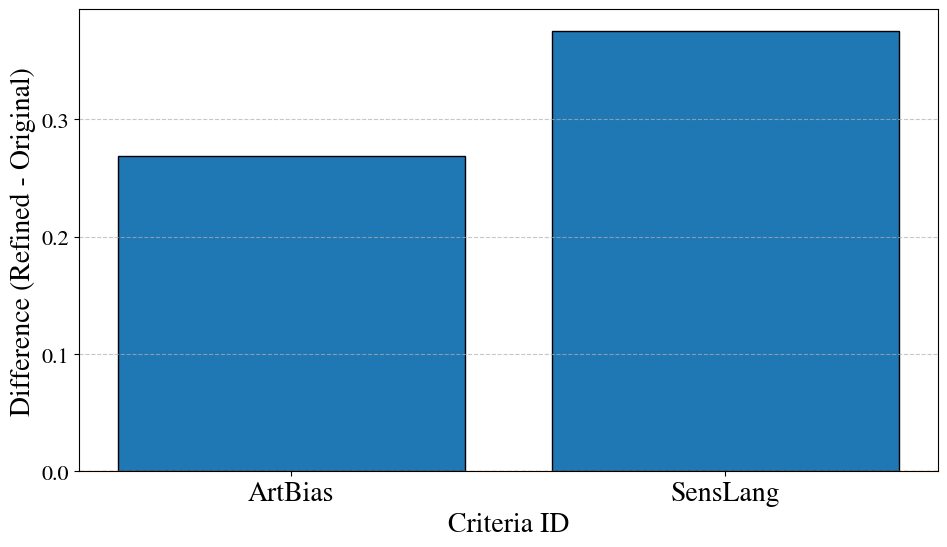} 
    \caption{Difference in agreement considering \textit{refined} and \textit{initial} criteria implementations. Only ArtBias and SensLang}\label{fig:agreement_comparison}
\end{figure}

Figure~\ref{fig:agreement_comparison} shows the gain in agreement when we use the refined prompt from Table~\ref{tab:gpt_adapted_selected_criteria_implementation} for the same two criteria, \textit{ArtBias} and \textit{SensLang}. The question remains unchanged, while the responses have become binary: Biased/Unbiased and Sensational/Neutral. 
What happens with the refined prompt is that \(k\) goes from 0.2064 to 0.4750 for \textit{ArtBias}  and from 0.1732 to 0.5486 for \textit{SensLang}.

\subsection{Resolution of  disagreements between human experts}\label{sec:disagreement_comparison}
In the previous section, we considered all the cases where human annotators agreed. We now focus on cases where annotators disagree, to see if LLM can help resolve disagreements.
We perform an ex-post human evaluation to create a ground truth, and cases where even this additional analysis does not provide a definitive answer are considered borderline.

We focus on the \textit{ArtBias, SensLang, NegTarg} and \textit{HeadAcc} criteria because these are the ones where the annotators disagreed the most. Furthermore, we only consider cases of relevant disagreement between annotators, i.e., cases where one annotator answered yes and the other did not, or, in the case of multiple answers, there are at least two degrees of separation between their answers (e.g., one says Inaccurate and the other says Quite Accurate).

Table \ref{tab:disagreement} shows, for the 5 criteria considered, (1) the total number of articles, (2) the number of articles where the human annotators gave different answers, (3) the number of times these disagreements are relevant, (4) the number of times the LLM gave the correct answer, and (5) the so-called borderline cases.

Articles where human annotators disagree on at least one criterion are 226. The table 
shows that, excluding borderline cases, the LLM resolves $100\%$ of disagreements for \textit{ArtBias} and \textit{NegTarg}, $81\%$ for \textit{HeadAcc}, and $63\%$ for \textit{SensLang}. 100\% for the two \textit{NegTarg} cases comes from the fact that LLM always gives a correct answer on non-borderline cases. 
The same is true for \textit{NegTarg} Identification.

\begin{table}[ht]
    \centering
    \small
    \caption{
    Statistics about disagreement between experts}
    \label{tab:disagreement}
    \begin{tabular}{ll}
    \hline
        \toprule
criteriaID & no\_articles / no\_art\_disagreement / \\&\textbf{relevant\_disagrements} / LLM-correct /\\& Borderline \\
\midrule
    ArtBias & 340 / 79 / \textbf{4}  / 4 / 0\\
    HeadAcc & 340 / 108 / \textbf{11} / 9 / 0\\
    NegTarg ( Detection ) & 340 / 30 / \textbf{30} / 18 / 12 \\
    NegTarg ( Identification ) & 340 / 47 / \textbf{47}  / 32 / 15\\
    SensLang & 340 / 72 / \textbf{11}  / 7 / 0 \\
        \bottomrule
    \end{tabular}
\end{table}

This highlights the LLM's potential to effectively assist humans in resolving disagreements related to the evaluation of these criteria.

For the degree-based criterion \textit{ArtBias}, the LLM resolves disagreements by selecting the central evaluation among those provided by human annotators. For example, if one annotator selected Biased and the other annotator selected Quite Unbiased, the resolution is with the annotation Quite Biased.

For \textit{NegTarg}, the LLM shows a strong ability to identify whether a group or individual is being negatively targeted within an article. However, there are several cases where even human experts find it difficult to make such determinations. The LLM shows similar proficiency when tasked with identifying the specific subject of negative targeting.

For \textit{SensLang}, although the number of relevant disagreements is limited, as in the cases of \textit{ArtBias}, we observe that the LLM is particularly sensitive to sensationalist content, especially content designed to evoke strong emotional responses in readers. For example, in nine cases where an annotator classified the content as "neutral", the LLM detected sensationalism levels later confirmed by the ground truth (corresponding to the ex-post human annotator).

\section{Discussion}\label{sec:discussion}
In this study, we investigated whether and to what extent one Large Language Model can automatically evaluate  criteria that journalistic organizations use in real-world evaluations to estimate reliability of online publishers. We adopt a subset of criteria used by the Global Disinformation Index (and very similar to some used by NewsGuard) to assess the risk of an online publisher exposing its readers to misinformation. 
The first of the research questions, \textbf{RQ1}, is whether it is possible to define quality prompts to help LLMs correctly evaluate good journalism criteria to assess the reliability level of a publisher.
The second of the research questions, \textbf{RQ2}, is how good is the LLM at aligning with the responses
given by human-experts, in the same context as before.
To answer these questions, we implemented an iterative refinement of the prompts. This process allowed us to improve some of the initial prompt definitions for better processing by the LLM. As described in Section \ref{sec:prompt_gain}, among the six criteria analyzed, the evaluation of three of them was already in good agreement with that of human experts. Specifically, in recognizing the nature of a news story, in recognizing the presence of a lede, and in understanding whether a text is negatively targeting something or someone. 
When it came to recognizing sensational or biased language, the initial prompts resulted in very low levels of agreement between the humans and the LLM. Upon deeper analysis of these results, we observe that LLMs show increased sensitivity in detecting bias and sensationalism compared to human experts. In addition, most errors are confined to neighboring classes, which mitigates the severity of prediction inaccuracies.
In particular, the use of refined prompts (offering binary responses instead of four options) significantly increases agreement, especially in detecting sensational language.


\textbf{RQ3} asks what types of questions LLMs can answer correctly and thus support the human annotator in the publisher's evaluation process.
To address this third question, we first note that in scenarios where increased annotation sensitivity is acceptable or a reduction in predictive granularity is warranted (e.g., considering binary instead of multiclass responses), the LLM-based annotator provides good quality annotations.
 As anticipated, this is particularly evident for criteria requiring significant effort, such as detecting the use of sensational language or assessing article bias in the text. Even better when it comes to judging whether the article is negatively targeting someone or something (for this criterion, \textit{NegTarg}, Detection, we got the highest \(k\) of 0.7089).

In addition, an LLM-based annotator can offer valuable assistance in resolving disagreements among human experts. Specifically, the LLM demonstrates promise in supporting decisions concerning \textit{ArtBias}, \textit{HeadAcc}, and \textit{NegTarg}. This includes detecting negatively targeted groups or individuals and accurately identifying them —tasks that are often complex and may not lead to definitive conclusions. In particular, accurate assessment of negative targeting of individuals or minority groups is critical to combating the spread of adversarial narratives. 
 \cite{gdiadversarial}.
The significant alignment achieved in a zero-shot experimental setting underscores the potential of the LLM for further improvement through tuning and adaptation. 

\section{Conclusions and Future Work}
We explored the use of one LLM to evaluate a set of criteria designed to assess online journalism standards. The LLM achieves substantial levels of agreement with human experts on complex evaluations, such as understanding whether the text negatively targets someone or something. In addition, we show that LLMs are highly effective at resolving disagreements among human experts in tasks such as detecting negative targeting and determining bias and headline accuracy in articles. 

As a future work we aim to expand the scope of our experiments in several key areas. First, we plan to expand the list of evaluation criteria to include aspects that consider visual elements, thus addressing a broader range of news article features. Second, we intend to increase the number of news outlets studied and articles  analyzed to improve the robustness and generalizability of our findings. In addition, we will explore the application of our methodology across multiple languages to assess its effectiveness in different linguistic contexts. 
Finally, we plan to experiment with a variety of LLM models and incorporate advanced techniques such as Retrieval-Augmented Generation (RAG) for prompt engineering and few-shot learning to further refine our evaluation framework.

\paragraph{Acknowledgements} 

This work is partially supported by SERICS (PE00000014) under the NRRP MUR program funded by the EU - \#NGEU. 

\bibliography{main}

\appendix

\section{Ethical Considerations}
Our research involves the use of LLMs to evaluate the reliability of online news publishers, a task which includes a step of annotating news articles according to journalistic criteria, usually done by human annotators.
While the ability of LLMs to assist in such evaluations offers some advantages, it is imperative to recognize and address the potential ethical implications of this technology.

\begin{itemize}
    \item Bias and Discrimination: Given that LLMs are trained on large datasets, they may inherently carry biases present in their training materials. In the context of news evaluation, such biases could affect the impartiality of trustworthiness assessments, potentially perpetuating stereotypes or unfairly targeting certain publishers based on biased training data.

    \item Transparency and Accountability: The decision-making process of LLMs is typically complex and not fully transparent, making it difficult to understand how or why certain scores are generated. This lack of clarity can be particularly problematic in scenarios where decisions need to be justified or contested.
    
    \item Environmental Impact: The computational demands of training and running LLMs are substantial, contributing to significant energy consumption and environmental impact. This is a critical consideration given the global push to reduce carbon footprints.

\end{itemize}

In using LLMs to support the evaluation of news publishers' reliability, it is important not only to harness their potential to improve efficiency and accuracy, but also to rigorously address these ethical concerns. By actively working to mitigate bias, increase transparency, and reduce environmental impact, we can better harness the power of LLMs in a way that is consistent with ethical standards and societal expectations.

\section{Online Resources}
Under request the authors will make the code and the data to replicate the experiments available. 
	
\end{document}